\documentclass[5p]{elsarticle}

\usepackage{lineno,hyperref,multirow,bm,tabularx}

\journal{Elsevier}

\begin{document}
\date{}
\begin{frontmatter}

\title{Spatial coherence characterization of light: an experimental study using digital micromirror devices}


\author[1,2]{Tiago E. C. Magalh{\~a}es\corref{mycorrespondingauthor}}
\cortext[mycorrespondingauthor]{Corresponding author}
\ead{tiago.e.magalhaes@alunos.fc.ul.pt}

\author[1,2]{Jos{\'e} M. Rebord{\~a}o}

\author[1,2]{Alexandre Cabral}

\address[1]{Departamento de F{\'i}sica, Faculdade de Ci{\^e}ncias, Universidade de Lisboa, Edif{\'i}cio C8, Campo Grande, PT1749-016 Lisboa, Portugal}
\address[2]{Instituto de Astrof{\'i}sica e Ci{\^e}ncias do Espa\c{c}o, Edif{\'i}cio C8, Campo Grande, PT1749-016 Lisboa, Portugal}

\begin{abstract}
We present spatial coherence measurements of partially coherent light in the far-field of incoherent sources with an experimental setup based on the Thompson-Wolf and Partanen-Turunen-Tervo experiments, to be performed in the context of a possible solar coherence measurement space instrument. The optical setup consists of a telescope to collimate light from a source, to diffract it by a digital micromirror device implementing a Young double-aperture interferometer in retroreflection mode, and finally to image the source into a two-dimensional sensor. Two multimode optical fibers with different diameters were used as incoherent sources and the results obtained for the spectral degree of coherence are compared to those expected from the van Cittert-Zernike theorem.
\end{abstract}

\begin{keyword}
Spatial coherence, Spatial light modulators, Statistical Optics
\end{keyword}

\end{frontmatter}


\section{Introduction}

The complex degree of coherence and the spectral degree of coherence are of critical importance to characterize spatial coherence in the space-time and space-frequency domains ~\cite{Wolf2007}, respectively. They are usually measured with Young experiments, i.e., double-aperture interferometers; by varying the separation between the apertures, it is possible to measure these quantities through an interference pattern in the observation plane. One pioneer work in this field was authored by Thompson and Wolf~\cite{Thompson1957} in 1957. In their experiment, light from an incoherent source propagates in free space and increases its spatial coherence (explained by the van Cittert-Zernike theorem) before impinging on a double-aperture. The modulation of the interference pattern is analyzed to retrieve the magnitude of the complex degree of coherence. However, to obtain a significant number of data points to estimate the complex degree of coherence, a large number of Young experiments must be performed with different separations between apertures and, in asymmetric cases, with different orientations (baselines) – a composite aperture must, therefore, be implemented to measure each data point of the complex degree of coherence. Since 1957, several experimental methods have been proposed (see~\cite{Divitt2014} and references therein) to measure spatial coherence. For example, Gonzalez and co-workers~\cite{Gonzalez2011} suggested using multi-aperture non-redundant arrays as an alternative to implementing simultaneously multiple Young experiments. Divitt et al.~\cite{Divitt2014} suggested using non-parallel slits, which have the advantage of measuring the spectral degree of coherence for broadband sources. In the same year, Partanen et al.~\cite{Partanen2014} used a digital micromirror device (DMD)~\cite{Hornbeck1997} to materialize the double-slit aperture to perform Young experiments for a multimode broad-area laser. Their method has the advantage of dynamically varying the separation between slits without having to implement several apertures and is only limited by dimensions and spatial resolution of the DMD. In 2017, experiments regarding the measurement of spatial coherence using DMDs for lensless imaging were reported~\cite{Kondakci2017,El-Halawany2017}. In this case, the authors used an incoherent light source in which light propagates a given distance until it is obstructed by an object. Spatial coherence of light diffracted by the object is measured using a DMD, and by applying back-propagation techniques, the object's size and location can be retrieved.

\begin{figure*}[t]
\centering
\includegraphics[width=0.90\textwidth]{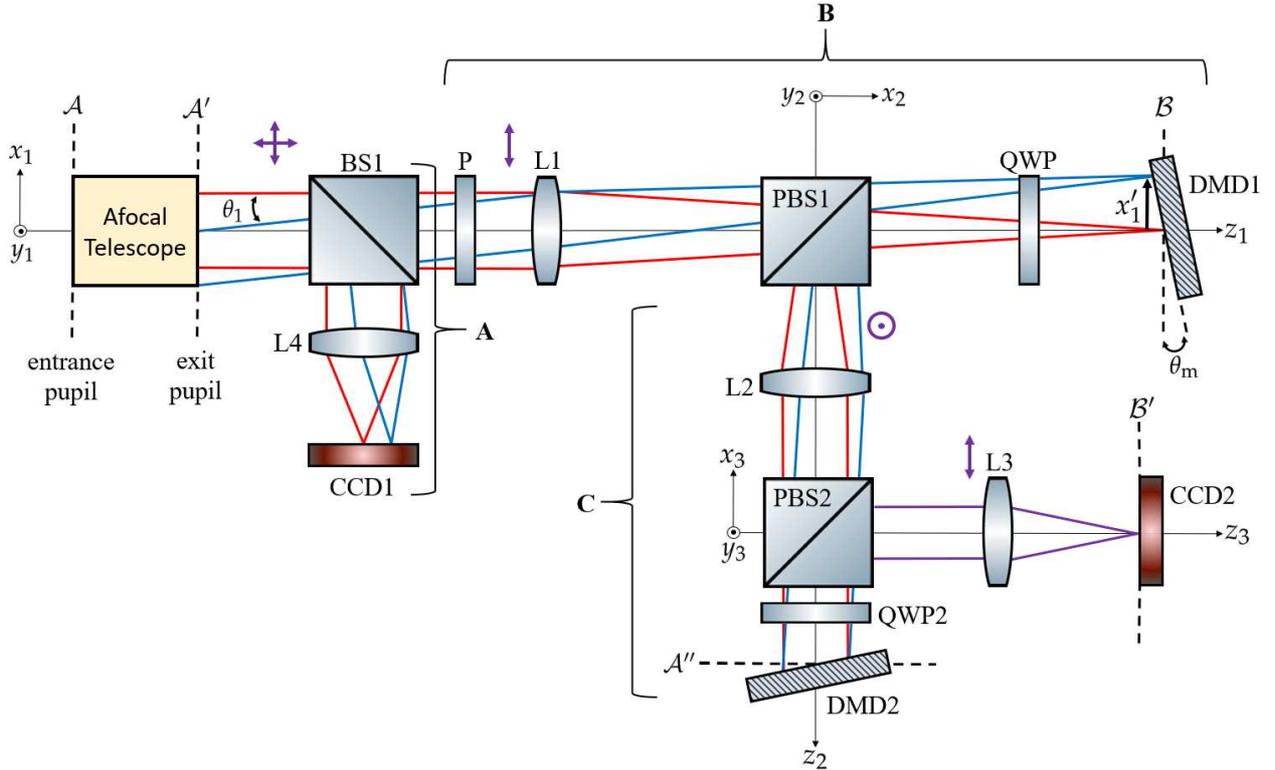}
\caption{(Color online) The Solar Coherence Instrument (SCI) optical setup (not to scale), described in~\cite{Magalhaes2019}. BS: Beam Splitter. PBS1, PBS2: Polarized Beam Splitters: QWP1, QWP2: Quarter-Wave Plates.
L1, L2, L3, and L4: Lenses. CCD1, CCD2: Charged Coupled Devices. DMD1, DMD2: Digital micromirror devices tilted by $\theta_m$. P: Polarizer. Polarization states are denoted by the double-ended arrows and by the circled dot. Each set of planes $\{ \mathcal{A}$ , $\mathcal{A}'$ , $\mathcal{A}''\}$  and $\{\mathcal{B}''$, $\mathcal{B}''\}$ correspond to conjugated planes. A, B, and C are the three subsystems of SCI. In this work, we validate experimentally subsystem C.}
\label{fig:sci}
\end{figure*}

Recently, a conceptual space-based instrument for the measurement of spatial coherence of structured light sources was proposed~\cite{Magalhaes2019} and its figure is replicated in Fig.~\ref{fig:sci}. 
In the design of this instrument, two DMDs are used in retroreflection mode to perform selective imaging and spatial coherence measurements. In this work, we experimentally verify this configuration regarding the spatial coherence measurements by using a DMD in retroreflection. The experimental apparatus is based on both the Thompson-Wolf~\cite{Thompson1957} and Partanen-Turunen-Tervo~\cite{Partanen2014} experiments. The light source is placed at the back focal plane of a positive lens to generate a collimated beam impinging on a DMD. The latter is used to perform Young double-slit experiments and is positioned to enable retroreflection of the diffracted light. The optical setup is compact and attractive for astronomical ground-based instruments that seek to measure the spatial coherence of light in the optical domain. Studies on the space robustness and operational stability of DMDs, through environmental testing~\cite{Fourspring2013,Travinsky2017} have been recently reported, anticipating their use in space. Our configuration is, in fact, envisaged for space-based optical instruments including balloon experiments. The conceptual instrument described in reference \cite{Magalhaes2019} is an example of such configuration.

In Section~\ref{sec:Setup} we present the experimental setup for measuring spatial coherence of light using a DMD in retroreflection mode. In Section~\ref{sec:Experiment} we validate this experimental setup by measuring spatial coherence of partially coherent light, where the spatial coherence function is described by the van Cittert-Zernike theorem. In Section~\ref{sec:Conclusions} we conclude and discuss future work.

\section{Description of the optical system}
\label{sec:Setup}

In this section, we start by introducing the key component of the optical setup, the DMD. We then describe the optical setup  and discuss the measurement of spatial coherence. An example of an incoherent circular source is presented, which will be useful later in Section~\ref{sec:Experiment}.

\subsection{Digital micromirror device}
The central component of the setup is a DMD in retroreflection mode, previously used for other applications~\cite{Riza2003,Park2004}. A DMD is a spatial light modulator, consisting of a $N_x\times N_y$ array of squared micromirrors with two stable angular orientations, tilted by $+\theta_m$  and $-\theta_m$ (of the order of $\sim15$ degrees), designated herein as "on" and "off" states, respectively. The rotation axis of individual micromirrors is usually along their diagonal. To perform retroreflection from the "on" state micromirrors, one has to tilt the DMD by $-\theta_m$ around the y-axis and by 45 degrees around the $z$-axis (see inset of Fig.~\ref{fig:setup}). Some DMDs do not require the latter rotation since micromirrors are already in-plane rotated by this amount. To the best of our knowledge, DMDs with a stable tilt angle of $\theta_m =0$ do not exist yet (in such cases, rotation of the DMD around the $y$-axis would not be necessary). However, other devices such as analog micromirror arrays can provide a zero degree tilt~\cite{Song2018}. For more information regarding DMDs and their applications, see~\cite{Ren2015}.

\subsection{Optical Setup}
In this paper, we will address an optical subsystem of a recently proposed instrument~\cite{Magalhaes2019} to measure the spatial coherence of structured sources, i.e., sources that are made up of small-scale sources (e.g., fiber bundles or the Sun's photosphere containing granular cells). The instrument is composed of three main subsystems (A, B and C) which are represented in Fig.~\ref{fig:sci}. Subsystem A is used for optical imaging. Subsystem B is used for selective imaging of small-scale sources. Subsystem C measures the spatial coherence of collimated light using a DMD in retroreflection mode. Our goal is to validate experimentally subsystem C.

\begin{figure}
\centering
\includegraphics[width=0.48\textwidth]{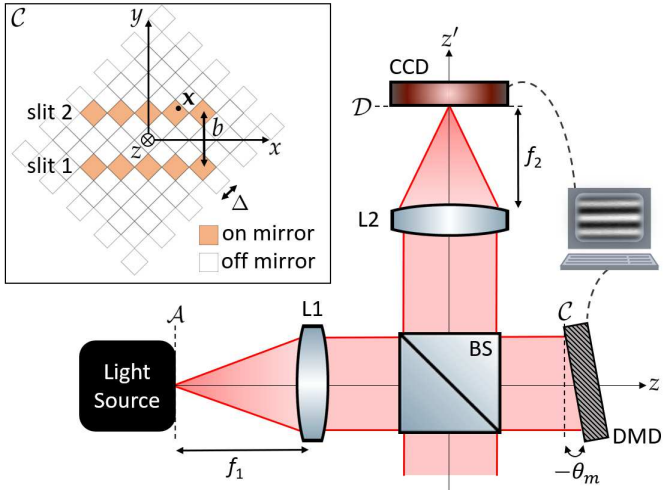}
\caption{(Color online) Schematic of the optical system for the spatial coherence measurements using a DMD in retroreflection (not to scale). L1, L2: Lenses. CCD: Charged Coupled Device. DMD: Digital Micromirror Device. BS: Beam Splitter. The inset shows the orientation and state of micromirrors in the DMD. $\Delta$ is the width of each micromirror.}
\label{fig:setup}
\end{figure}

The layout of the optical setup is depicted in Fig.~\ref{fig:setup}. A light source with central angular frequency $\omega_0$ is collimated by lens L1 with focal length $f_1$. Light passes through a beam splitter (BS) and reaches the DMD plane, $\mathcal{C}$. The incoming reflected beam from the BS is lost out of the system, although in other instruments (e.g., afocal telescopes), this beam can be used for other purposes, such as complementary imaging and spectroscopy. With the "on" state mirrors of the DMD, a Young double-slit aperture with separation $b$ can be implemented, as represented in the inset of Fig.~\ref{fig:setup}. Light reflected from the "off" state mirrors is, again, lost out of the system. The DMD is tilted around the $y$-axis by $\theta_m$ degrees to ensure retroreflection for the "on" state mirrors. In case the micromirrors are not in-plane rotated, one has to tilt the DMD along the $z$-axis by 45 degrees. After retroreflection, light is reflected in the BS and passes through lens L2 with focal length $f_2$ which creates a magnified image of the source at the CCD plane $\mathcal{D}$. Both the DMD and the CCD are connected to a computer that dynamically changes the separation of the Young slits in the DMD and acquires and analyzes the interference patterns to retrieve the spatial coherence function. We will discuss next the Young double-aperture experiment used to achieve this goal.

\subsection{Spatial coherence measurement}

Young double-aperture experiments are performed in plane $\mathcal{C}$ using a DMD - the relevant theory is described elsewhere~\cite{Magalhaes2019}. The quantity of interest is the complex degree of coherence $\gamma(\mathbf{x}_1,\mathbf{x}_2,\tau)$ or the spectral degree of coherence $\mu(\mathbf{x}_1,\mathbf{x}_2,\omega)$, where $\mathbf{x}_1$ and $\mathbf{x}_2$ are positions at apertures 1 and 2, respectively, $\tau$ is the time difference in a Young's experiment~\cite{Wolf2007} and $\omega$ is the angular frequency.

Two assumptions are going to be made. First, we assume that light has the same normalized spectrum in both apertures, i.e.,
\begin{equation}
    s_1(\omega)=s_2(\omega)=s(\omega)\,,\label{eq:cond1}
\end{equation}
where $s_1(\omega)$ and $s_2(\omega)$ are the normalized spectra at aperture 1 and 2, respectively. Second, we also assume that light is quasi-monochromatic, i.e.,
\begin{equation}
    \Delta\omega \ll \omega_0\,,\label{eq:cond2}
\end{equation}
where $\omega_0$ central frequency and $\Delta\omega$ the effective bandwidth range. It can therefore be proved that the spectral degree of coherence satisfy~\cite{Friberg1995}:
\begin{equation}
    \gamma(\mathbf{x}_1,\mathbf{x}_2,\tau) = \gamma(\mathbf{x}_1,\mathbf{x}_2,0)\,\exp\left(-\mathrm{i}\omega_0\tau\right)\,,\label{eq:t1}
\end{equation}
\begin{equation}
    \gamma(\mathbf{x}_1,\mathbf{x}_2,\tau) = \mu(\mathbf{x}_1,\mathbf{x}_2,\omega_0)\,\exp\left(-\mathrm{i}\omega_0\tau\right)\,.\label{eq:t2}
\end{equation}
From equation~(\ref{eq:t1}) we can conclude that temporal coherence will not affect the magnitude of the complex degree of coherence since $|\gamma(\mathbf{x}_1,\mathbf{x}_2,\tau)|=|\gamma(\mathbf{x}_1,\mathbf{x}_2,0)|$. From Equation~(\ref{eq:t2}) we can conclude that $|\gamma(\mathbf{x}_1,\mathbf{x}_2,\tau)| = |\mu(\mathbf{x}_1,\mathbf{x}_2,\omega_0)|$.

Assuming that light impinging in the DMD fulfills conditions~(\ref{eq:cond1}) and~(\ref{eq:cond2}), the intensity $I(\mathbf{u})$ measured by the CCD at plane $\mathcal{D}$ is given by~\cite{Wolf2007,Magalhaes2019}
\begin{eqnarray} 
I(\mathbf{u})	& = &	I_1(\mathbf{u})+I_2(\mathbf{u})+2\sqrt{I_1(\mathbf{u})}\sqrt{I_2(\mathbf{u})} \nonumber \\
&  &
		\times\left|\mu_{\mathcal{C}}(b,\omega_0)\right|\cos\left[\beta(b,\omega_0)-\frac{\omega_0 b v}{c f_2}\right]
 \,,\label{eq:I(u)}
\end{eqnarray} 
where $\mathbf{u}=(u,v)$, $I_1(\mathbf{u})$ and $I_2(\mathbf{u})$ are the observed intensities individually due to slit 1 and 2, respectively, $c$ is the speed of light, $\mu_{\mathcal{C}}(b,\omega_0)$ is the spectral degree of coherence at plane $\mathcal{C},$ $\beta(b,\omega_0)$ is the phase of the spectral degree of coherence \cite{Wolf2007}, i.e.,
\begin{equation}
    \mu_{\mathcal{C}}(b,\omega_0) = |\mu_{\mathcal{C}}(b,\omega_0)| \mathrm{e}^{ \mathrm{i}\beta (b,\omega_0) }\,,
\end{equation}
where $b$, the center-to-center separation between the slits (baseline), is given by
\begin{eqnarray}
    b=\left| \mathbf{x}_1-\mathbf{x}_2\right|\,,
\end{eqnarray}
where $\mathbf{x}=(x,y)$ (see inset of Fig.~\ref{fig:setup}). If light reaching the DMD has a non-negligible degree of spatial coherence for a given separation $b$ (i.e., $|\mu(b,\omega_0)|>0$), an interference pattern is observed in the CCD, provided three conditions are fulfilled: (1) the CCD has the adequate spatial resolution to resolve the interference fringes, (2) the period of the fringe pattern is smaller than CCD dimensions and (3) the signal-to-noise ratio (SNR) is sufficient to quantify the amplitude of the fringes. Lenses have a direct impact on the results because L2 determines the period of the fringes in the CCD and L1 is responsible for the observed range of the spectral degree of coherence in plane $\mathcal{C}$.

\subsection{Example of circular incoherent source}

We will now briefly discuss the propagation of light of an incoherent circular source between planes $\mathcal{A}$ and $\mathcal{C}$ - these results will be useful to analyze the experiments using this type of source in the next section. We will use the space-frequency approach, i.e., spatial coherence will be characterized by the spectral degree of coherence instead of the complex degree of coherence.

\begin{figure}[htb]
\centering
\includegraphics[width=0.49\textwidth]{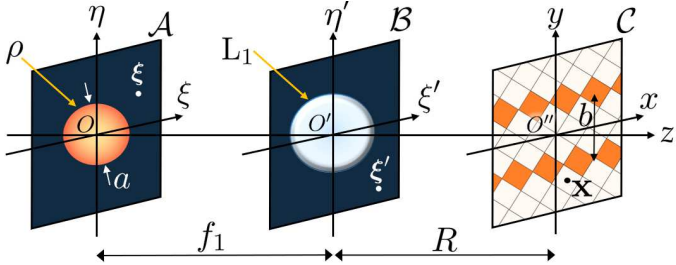}
\caption{(Color online) Notation used for the propagation of spatial coherence from plane $\mathcal{A}$ to $\mathcal{C}$ (see Fig. \ref{fig:setup}). $a$ is the diameter of source $\rho$ and $R$ is the distance between planes $\mathcal{B}$ and $\mathcal{C}$. $O$, $O'$ and $O'$ are the origins of planes $\mathcal{A}$, $\mathcal{B}$, and $\mathcal{C}$, respectively. The darker squares in plane $\mathcal{C}$ represent the "on" state micromirrors and $b$ is the separation between slits. }
\label{fig:theory}
\end{figure}

Let $\rho$ be a perfectly incoherent, quasi-monochromatic, and circular light source with central frequency $\omega_0$ and diameter $a$, in plane $\mathcal{A}$, as illustrated in Fig.~\ref{fig:theory}, which can represent, for instance, the finite ending tip of an optical fiber. A positive lens L1 with focal length $f_1$ is placed a distance $f_1$ from plane $\mathcal{A}$. The propagation of light from $\mathcal{A}$ to $\mathcal{B}$ is described by the van Cittert-Zernike theorem~\cite{MandelWolf1995}. Assuming that the physical extent of light impinging on lens L1 is smaller than the lens aperture, we can neglect the pupil function and, therefore, diffraction effects due to the lens finite-size aperture. Since the light source is at the back focal plane of L1, the magnitude of the spectral degree of coherence $\mu_{\mathcal{C}}(b,\omega_{0})$ at plane $\mathcal{C}$  will be the same as in plane $\mathcal{B}$ \cite{Wolf2007,Goodman2015} and is given by
\begin{equation}
\left|\mu_{\mathcal{C}}\left(b,\omega_{0}\right)\right| = 2\left|\frac{\mathrm{J}_{1}\left[\omega_0\, a\,b/\left(2cf_{1}\right)\right]}{\omega_0\, a\,b/\left(2cf_{1}\right)}\right|\,.
\label{eq:SDC_C}
\end{equation}
where $J_1$ is the first-order Bessel function of the first kind. The factor $a/f_1$ in Eq.~(\ref{eq:SDC_C}) will determine the range of spatial coherence in plane $\mathcal{C}$. To quantify the spatial coherence of light, we can define the effective correlation length $\sigma_{\mu}$  as the first zero of the first-order Bessel function:
\begin{equation}
\sigma_{\mu}\approx 7.66\frac{c\,f_1}{\omega_0\, a} \,. \label{eq:correlation_length}
\end{equation}
With the focal length $f_1$ for and the diameter $a$ of the light source, one can estimate the number of data points that can be acquired within the central lobe of the Bessel function.

\section{Experimental Validation}
\label{sec:Experiment}

We will experimentally validate the optical setup described in Section~\ref{sec:Setup} by measuring the spectral degree of coherence created by an incoherent circular source. We use a spatially incoherent LED source (Thorlabs M660FP1) with central wavelength $\lambda_0=660\,\mathrm{nm}$ and a full width at half maximum of $\Delta\lambda =18\,\mathrm{nm}$ that is coupled to a multimode optical fiber. Since $\Delta\lambda\ll\lambda_0$, we can consider light to be quasi-monochromatic (see, for example, reference~\cite{BornWolf1998}). Since the LED source is spatially incoherent, the region at the end of a multimode fiber, placed in plane $\mathcal{A}$ (see Figs.~\ref{fig:setup} and~\ref{fig:theory}), can be modeled as an incoherent circular planar source~\cite{yoshimura1992}.

We also use the Texas Instruments DLP7000 DMD model; micromirrors are not rotated in the $x-y$ plane (see the inset of Fig.~\ref{fig:setup}) and, therefore, a 45 degrees tilt is needed to ensure retroreflection. The DMD has $1024\times768$ square micromirrors (width of $\Delta=13.68\,\mu\mathrm{m}$) and $\theta_m =12$ degrees. The CCD is AVT Marlin F-080, with  $1032\times778$ pixels and a pitch of $4.65\,\mu\mathrm{m}$. 

\begin{figure}[t]
\centering
\includegraphics[width=0.47\textwidth]{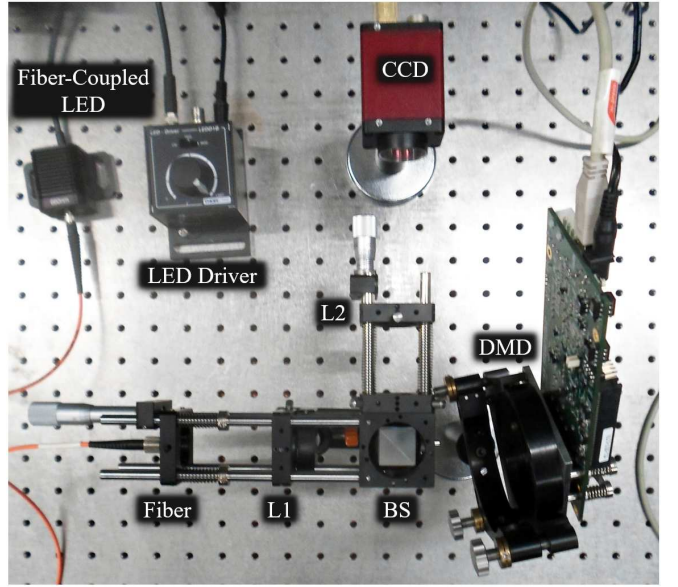}
\caption{(Color online) A photograph of the apparatus used to measure the spatial coherence of light at the DMD plane of fiber OF2. Abbreviations are the same used in Fig.~\ref{fig:setup}.}
\label{fig:photo}
\end{figure}

We performed experiments with two different multimode fibers, OF1 and OF2, with diameters $200\,\mu\mathrm{m}$ (Thorlabs M75L01) and $50\,\mu\mathrm{m}$  (Thorlabs M16L01), respectively. For OF1, the focal length of lenses L1 and L2 are $f_1=60\,\mathrm{mm}$ and $f_2=250\,\mathrm{mm}$, respectively, while for OF2, $f_1=60\,\mathrm{mm}$ and $f_2=100\,\mathrm{mm}$. A photograph of the setup is shown in Fig.~\ref{fig:photo} for OF2. The estimated effective correlation lengths for OF1 and OF2 are $\sigma_\mu=0.241\,\mathrm{mm}$ and $\sigma_\mu=0.966\,\mathrm{mm}$, respectively. Therefore, a larger number of data points within the central lobe of the Bessel function will be available for OF2 since the effective correlation length is larger.

For each separation $b$, the digital intensity pattern $I(\mathbf{u})$ is represented by a two-dimensional array. To extract the spectral degree of coherence from Eq.~(\ref{eq:I(u)}), we measured the intensity pattern of the double-slit, $I(\mathbf{u})$, and also the patterns from each slit, $I_1(\mathbf{u})$ and $I_2(\mathbf{u})$, and defined a new array $I_M(\mathbf{u})$~\cite{Partanen2014}:
\begin{eqnarray} 
I_M(\mathbf{u}) & = &\frac{I(\mathbf{u})-I_1(\mathbf{u})-I_2(\mathbf{u})}{2\sqrt{I_1(\mathbf{u})}\sqrt{I_2(\mathbf{u})}} \nonumber \\
& = &
  \left|\mu_{\mathcal{C}}(b,\omega_0)\right|\cos\left[\beta(b,\omega_0)-\frac{\omega_0 b v}{c f_2}\right]\,.\label{eq:Il(u)}
\end{eqnarray}
The intensity pattern with all micromirrors in the "off" state is also acquired and subtracted from the other measurements. In this case, we noticed that a very faint pattern was observed in the CCD, possibly emerging from the tiny linear separation between micromirrors and, therefore, diffraction.  To increase the SNR, we increased the integration time and the gain of the CCD. However, by increasing the gain, we noticed that the number of saturated pixels increased, and such pixels had to be discarded. We performed a binning process to further increase the SNR, which consists of combining pixels of the CCD in both horizontal and vertical directions. However, vertical binning reduces spatial resolution, which could be critical to retrieve the higher spatial frequencies of the spectral degree of coherence. With regard to horizontal binning, one must be careful since some columns may be noisier than others. The values of $I_M(\mathbf{u})$ are fitted using the following template:
\begin{equation}
    g_1(y',b,A,B,C) = A\cos\left[\frac{\omega_0\,b\,(y'-B)}{c\,f_2}\right]+C\,, \label{eq:fit_func}
\end{equation}
where $A$ corresponds to the magnitude of the spectral degree of coherence $\mu_{\mathcal{C}}(b,\omega_0)$, for a given $b$, $B$ corresponds to the shift (dephasing) of the center of the diffraction pattern and $C$ accounts for background noise and high gain effects on the CCD. $B$ is also determined using interferograms. Nonetheless, in the fitting process, we allow for minor corrections, even though it is well constrained to a small set of CCD pixels. For OF2, a nonlinear chirp term was added, since the frequency of the fringe pattern was not constant throughout $y'$ (most probably due to residual lens distortion). The modified fitting function is
\begin{eqnarray}
    g_{2}(y',b,A,B,C) & = & A\cos\left\{ \frac{\omega_0\, b}{c\, f_{2}}\left[\left(y'-B\right)\right.\right.\nonumber \\
 &  & +\left.\left.D\left(y'-B\right)^{2}\right]\right\} +C\,,
\label{eq:fit_func2}
\end{eqnarray}
where $D$ accounts for the nonlinear chirp. Note that $D$ is treated as a constant and not a variable of function $g_2$, since it does not depend on the separation $b$.
\begin{figure}[t]
\centering
\includegraphics[width=0.49\textwidth]{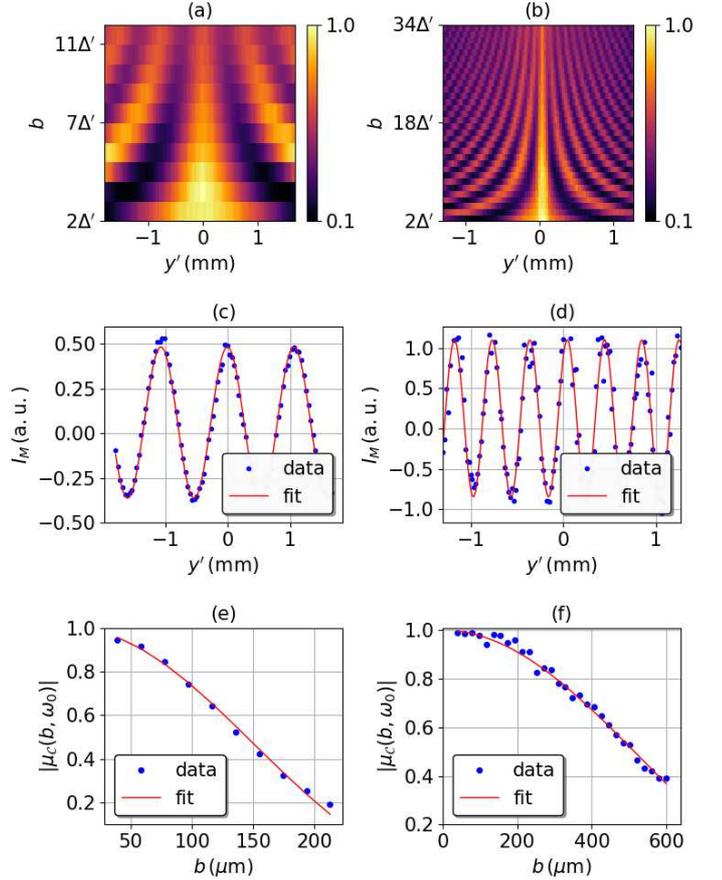}
\caption{(Color online) Spatial coherence results for fibers OF1 and OF2. (a), (c), and (e) correspond to OF1, while (b), (d), and (f) corresponds to OF2. (a) and (b) are normalized intensity interferograms. (c) and (d) correspond to $I_M(v)$ for $b=8\Delta'$, where $\Delta'=\sqrt{2}\Delta$ and $\Delta$ is the micromirror size ($\Delta=13.68\,\mathrm{\mu m}$).  (e) and (f) are the results for the magnitude of the spectral degree of coherence at plane $\mathcal{C}$.}
\label{fig:results}
\end{figure}

The results for both fibers are presented in Fig.~\ref{fig:results}.
Intensity interferograms in Figs.~\ref{fig:results}(a) and~\ref{fig:results}(b) have different frequencies for different separation between slits, $b$, as expected from Eq.~(\ref{eq:I(u)}). A sample of results for $I_M(v,b)$ for both fibers are represented in Figs.~\ref{fig:results}(c) and~\ref{fig:results}(d). The fit used for OF1 corresponds to the function $g_1$ of Eq.~(\ref{eq:fit_func}), while for OF2 corresponds to $g_2$ of Eq.~(\ref{eq:fit_func2}). The magnitude of the spectral degree of coherence obtained from the previous fittings for each separation $b$ is represented in Figs.~\ref{fig:results}(e) and~\ref{fig:results}(f) for fibers OF1 and OF2, respectively. For fiber OF2, the estimated value for the nonlinear chirp was $D=4.64\,\mathrm{m}^{-1}$. The data from the magnitude of the spectral degree of coherence $|\mu_{\mathcal{C}}(b,\omega_0)|$ was fitted using Eq.~(\ref{eq:SDC_C}) in order to retrieve the diameter $a_1$ and $a_2$ of fibers OF1 and OF2, respectively. The values obtained were $a_1=192\pm2\,\mu\mathrm{m}$ and $a_2=54.3\pm0.6\,\mu\mathrm{m}$, showing deviations of $4\%$ and $9\%$, respectively, with respect to manufacturers' specifications. Nevertheless, these estimations of the diameter of the fibers do assume that the source is incoherent and that the theorem of van-Cittert-Zernike fully applies. The results are summarized in Table~\ref{tab:results}.
\begin{table}
\protect\caption{Summary of results. The retrieved values for the diameters of the fibers do match well with values given by the manufacturer.\label{tab:results}}
\centering{}%
\begin{tabular}{|c|>{\centering}m{0.16\linewidth}|>{\centering}p{0.16\linewidth}|>{\centering}m{0.16\linewidth}|c|}
\hline 
\multirow{2}{*}{Fibers} & \multicolumn{2}{c|}{Focal length $(\mathrm{mm})$} & \multicolumn{2}{c|}{Diameter $(\mu\mathrm{m})$}\tabularnewline
\cline{2-5} 
 & $f_{1}$ & $f_{2}$ & manuf. spec. & retrieved\tabularnewline
\hline 
\hline 
OF1 & 60 & 250 & 200 & 192\tabularnewline
\hline 
OF2 & 60 & 100 & 50 & 54\tabularnewline
\hline 
\end{tabular}
\end{table}

Spatial coherence measurements may take some minutes. The time depends on the number of interferograms and on the integration time of the CCD. If the light source intensity changes significantly while performing the measurement, the spectral degree of coherence cannot be retrieved correctly, which could be a problem to study some light sources. However, it is worthwhile studying new patterns of micromirrors on the DMD such as non-parallel slit by Divitt et al.~\cite{Divitt2014}, instead of the double-slit pattern, in order to retrieve the spectral degree of coherence faster and possibly with the adequate spectral resolution for broadband sources. Another problem that can affect spatial coherence measurement is the tilting of the DMD around the $y$-axis by $-\theta_m$. This means that there will be different path lengths from opposite edges of the DMD, as pointed out by Partanen et al.~\cite{Partanen2014}. In our case, since light is collimated, the difference in spatial coherence is not significant for the different edges of the DMD. Nevertheless, it still affects imaging and these phase effects are worth studying. 

\section{Conclusions}
\label{sec:Conclusions}
The magnitude of the spectral degree of coherence of partially coherent light generated by an incoherent circular source was measured using a setup based on the Thompson-Wolf and Partanen-Turunen-Tervo experiments with a digital micromirror device acting as a Young double-aperture interferometer in retroreflection mode. Light from a fiber tip, coupled to a LED, was used as a light source. The van Cittert-Zernike theorem was used to model the propagation of partially coherent light and the values for the spatial coherence were compared to the theoretical ones. 

To use this setup integrated with a payload of ground and space-based instruments, one would have to replace the fiber by an imaging lens in order have a secondary light source in plane $\mathcal{A}$. Assuming that the imaged light source is incoherent, one would select lens L1 according to the spatial dimensions and distance to the source, using the van-Cittert Zernike theorem as a reference. Source reconstruction from spatial coherence measurements is an example of what these measurements can provide, as mentioned in Ref.~\cite{Beckus2018}. We also point out that, by using selective imaging, one could image, in plane $\mathcal{A}$, a region of interest to perform spatial coherence measurements. This is the basis of the conceptual instrument proposed recently~\cite{Magalhaes2019}. Nonetheless, further studies are needed to evaluate phase problems when measuring spatial coherence with DMDs of light sources not modeled by the van Cittert-Zernike theorem, i.e., partially coherent sources. Finite-size aperture effects~\cite{thompson1984finite} may also have to be assessed whenever these effects are of the same order of magnitude of the effective correlation length of light at plane $\mathcal{B}$ for a given separation $b$.

\section*{Acknowledgment}
Tiago E. C. Magalh{\~a}es acknowledges the support from Fun\-da\-\c{c}{\~a}o para a Ci\-{\^e}n\-cia e a Tec\-no\-lo\-gia (FCT, Portugal) through the grant PD/BD/105952/2015, under the FCT PD Program PhD::SPACE (PD/00040/2012). This work was supported by FCT/MCTES through national funds (PIDDAC) by grants UID/FIS/04434/2019 and UIDB/04434/2020.

\bibliography{PhDSpace_Ref}

\end{document}